\documentclass[a4paper,12pt]{article}
\usepackage{amssymb, amsmath, hyperref}
\usepackage[final]{graphicx}
\usepackage[latin1]{inputenc}

\begin{document}

\title{A graviton propagator for inflation}
\author{Tomas Janssen\footnote{T.Janssen2@phys.uu.nl} and Tomislav Prokopec\footnote{T.Prokopec@phys.uu.nl}\\ \emph{Institute for Theoretical Physics, University of
Utrecht}} \maketitle  \flushright{ITP-UU-07/37, SPIN-07/25}
\flushleft  \abstract{We construct the scalar and graviton
propagator in quasi de Sitter space up to first order in the slow
roll parameter $\epsilon\equiv -\dot{H}/H^2$. After a rescaling,
the propagators are similar to those in de Sitter space with an
$\epsilon$ correction to the effective mass. The limit
$\epsilon\rightarrow 0$ corresponds to the $E(3)$ vacuum that
breaks de Sitter symmetry, but does not break spatial isotropy and
homogeneity. The new propagators allow for a self-consistent,
dynamical study of quantum back-reaction effects during
inflation.}

\section{Introduction}
While inflation --the accelerated expansion of the early
universe-- was originally introduced to solve the horizon,
flatness and cosmic relic
problem~\cite{Guth:1980zm}\cite{Liddle:1999mq}, it was soon
realized that inflationary space-times have the interesting
property of particle
creation~\cite{Mukhanov:1981xt}~\cite{Starobinsky:1979ty}~\cite{Birrell:1982ix}.
This can be seen from a simple application of the energy-time
uncertainty relation~\cite{Prokopec:2003bx}\cite{Woodard:2003vp}
that states that a virtual particle-anti particle pair with energy
$E$ and momenta $\pm k$ can exist a time $\delta t$ given by
\begin{equation} \label{uncertainty}
    \int_t^{t+\delta t} dt' E(t',k)\lesssim 1.
\end{equation}
While in flat space $\delta t$ is very small, it can grow large in
curved space-time. In particular in the context of cosmological
inflation, one finds that for massless modes with wavelengths
larger then the hubble radius, $\delta t$ can become infinite.
This means that during inflation massless, infrared particles are
created out of the vacuum. Since photons and massless fermions
couple conformally, their production rate is suppressed by the
scale factor. The only non-conformally coupled, massless particles
are massless minimally coupled scalars and gravitons.\\
Based on the discussion above one must conclude that quantum
effects from massless minimally coupled scalars and gravitons
during inflation can be very substantial. Indeed, these effects
are exactly the source for scalar and tensor cosmological
perturbations~\cite{Mukhanov:1981xt}. The imprint scalar
perturbations leave on the cosmic microwave background has been
verified by WMAP~\cite{Spergel:2006hy}. Tensor perturbations might
be detected by the Planck satellite.\\
In general doing quantum field theory on a curved background is
very involving, but, due to its high amount of symmetry, much more
can be done in the context of de Sitter space (section
\ref{sdesit}). Since in de Sitter space the scale factor grows
exponentially with time, it is an excellent paradigm for an
inflationary space. The $D$-dimensional propagators in de Sitter
space are
known\cite{Chernikov:1968zm}~\cite{Onemli:2002hr}\cite{Woodard:2004ut}\cite{Woodard:2005cw}
and therefore it is possible to use dimensional regularization to
calculate many kinds of interesting and possibly observable
quantum effects during inflation. For example a massless minimally
coupled scalar with a quartic self-interaction has been
studied~\cite{Onemli:2002hr}\cite{Onemli:2004mb}\cite{Brunier:2004sb}.
The resulting model shows a violation of the weak energy condition
on cosmological scales. A non-minimally coupled, massive scalar
with quartic self-interaction has also been studied
\cite{Bilandzic:2007nb}. In this model the radiative corrections
to slow roll inflation where calculated and found to be
unobservable.\\
Also scalar electrodynamics has been studied
extensively\cite{Prokopec:2003bx}\cite{Prokopec:2002jn}\cite{Prokopec:2002uw}\cite{Prokopec:2003iu}\cite{Prokopec:2006ue}\cite{Dimopoulos:2001wx}.
The vacuum polarization has been studied and it has been shown
that in such a model superhubble photons acquire a mass. The
contribution to the zero-point energy of these photons sources
cosmological
magnetic fields. Most recently the theory has been formulated stochastically~\cite{Prokopec:2007ak}.\\
 Another interesting line of
research is the study of quantum back-reaction on the
vacuum\cite{Tsamis:1996qm}\cite{Tsamis:1996qq}\cite{Abramo:1996gd}\cite{Abramo:1997hu}\cite{Losic:2006ht}\cite{Janssen:2007yu}\cite{Abramo:2001dd}.
In a pioneering study~\cite{Tsamis:1996qm}\cite{Tsamis:1996qq}, it
was shown that the two loop contribution to the graviton one-point
function slows inflation. The resulting model, where inflation is
driven by a large cosmological constant and ends naturally due to
quantum effects is an alternative to scalar field driven
inflation. Moreover it might prove to be important for a better
understanding of the cosmological constant
problem\cite{Tsamis:1992sx}. However this calculation was done in
$3+1$ dimensions, using a (time dependent) ultraviolet momentum
cutoff. Since it is not clear whether the observed effect is a
relic of the regularization procedure, it is important that
this calculation is redone using dimensional regularization\cite{Tsamis:2005je}.\\
There are however several drawbacks to the use of de Sitter space.
First of all, while de Sitter space provides an excellent
framework to study properties of inflation, it is never truly
realized in nature. Indeed, since in de Sitter space the Hubble
parameter is per definition globally constant, it follows that if
the universe was once de Sitter it is always de Sitter. From the
fact that the universe is not de Sitter today, it follows that it
was never de Sitter. The deviation of the inflationary universe
from de Sitter space is even measurable through the deviation of
the spectral index $n_s$ from scale invariance
\cite{Liddle:1999mq}\cite{Spergel:2006hy}. When these deviations
are small enough, one could work in a space that is 'locally' de
Sitter, however in that case there is no control on the error one is making. \\
 The second problem with de Sitter space is that it is
non-dynamical. While space-time is certainly curved, all dynamics
are governed by the Hubble parameter and the Hubble parameter is
constant. Strictly speaking it is therefore inconsistent to study
how space-time changes due to back-reaction effects if the
calculation of these effects explicitly assumes that the
background is constant. To self-consistently consider
back-reaction, one must allow the Hubble parameter to change with
time. Although in~\cite{Tsamis:1996qm}\cite{Tsamis:1996qq} a
distinction is made between the constant Hubble parameter $H$ and
the -- dynamical-- physical observable $H_{\rm eff}$, it is still
not fully consistent. The reason is that the propagators are
calculated using the constant $H$, where they should be calculated
with the dynamical, time dependent $H_{\rm eff}$.\\
 The final issue
with de Sitter space is the fact that it is not possible to
construct a propagator for a massless minimally coupled scalar or
for the graviton that respects the de Sitter
symmetry\cite{Allen:1985ux}. Usually this problem is circumvented
by letting these propagators break the de Sitter symmetry, but not
spatial isotropy and homogeneity. The argument is that it is these
symmetries one expects to be present in the early universe, not
the full de Sitter group (because of the problem described above).
However, if we are forced to abandon de Sitter symmetry anyway,
why not do the complete calculation in a space with only spatial
isotropy and homogeneity, but otherwise
similar to de Sitter space.\\
Constructing the propagator in this space, known as quasi de
Sitter space, is exactly the aim of this paper. Quasi de Sitter
space is a generalization of de Sitter space, where $H$ depends
only mildly on time (see section \ref{sQDS}). While this time
dependence breaks the full de Sitter symmetry, it preserves
spatial homogeneity and isotropy. We therefore find that there is
no problem to describe massless minimally coupled scalars or
gravitons in quasi de Sitter space. Moreover, since quasi de
Sitter space is both dynamical and a more realistic approximation
in many inflationary
models then de Sitter space, the other two problems mentioned above are also resolved.\\

In section \ref{sdesit} we overview some properties of de Sitter
and quasi de Sitter space-times. In sections \ref{sscal} and
\ref{sgrav} we calculate the scalar and graviton propagators
respectively. In section \ref{sgrav} we also calculate the ghost
propagator associated with our gauge fixing. We conclude in
section \ref{scon}.

\section{de Sitter and quasi de Sitter space}\label{sdesit}
4-dimensional de Sitter space is the hypersurface given by
\begin{equation}
    -X_0^2+X_1^2+X_2^2+X_3^2+X_4^2=\frac{1}{H^2}
\end{equation}
embedded in 5-dimensional Minkowski space-time, where $H$ is a
constant parameter known as the Hubble parameter. de Sitter space
is a solution to the Einstein equations with a positive
cosmological constant, in which case
\begin{equation}
    H^2=\frac{\Lambda}{3}.
\end{equation}
The isometry group of de Sitter space, $SO(1,4)$, is manifest in
this embedding. We will use flat coordinates, which cover only
half of the de Sitter manifold, given by ($i=1,2,3$)
\begin{equation}
    \begin{split}
        X_0&=\frac{1}{H}\sinh(Ht)+\frac{H}{2}x_i x^i e^{H t},\\
        X_i&=e^{H t}x_i,\\
        X_4&=\frac{1}{H}\cosh(Ht)-\frac{H}{2}x_i x^i e^{H t},\\
        -\infty&<t,x_i<\infty.
    \end{split}
\label{coordinates}
\end{equation}
In these coordinates the metric reads
\begin{equation}\label{dsmetric}
    g_{\mu\nu}=\text{diag}(-1,a^2,a^2,a^2)\qquad\qquad a=e^{H t}
\,,
\end{equation}
such that $\dot{a}/a=H$. The coordinates (\ref{conformal}) can be
written in conformal form by changing coordinates to conformal
time $\eta$ defined as $ad\eta=dt$. The metric becomes:
\begin{equation}
    g_{\mu\nu}=a^2\eta_{\mu\nu},\qquad
    a=-\frac{1}{H\eta},\qquad\eta_{\mu\nu}=\text{diag}(-1,1,1,1)
, \qquad(\eta<0) \,.
 \label{conformal}
\end{equation}

 We define the de Sitter invariant distance functions
\begin{equation}\label{dsfunctions}
Z(X;\tilde{X})=H^2\eta_{AB}X^A\tilde{X}^{B}
=1-\frac{1}{2}Y(X;\tilde{X}) \,.
\end{equation}
In conformal coordinates these functions read
\begin{equation} \label{QDS_y}
    \begin{split}
        z_{++}(x;\tilde{x})&=1-\frac{1}{2}y_{++}(x;\tilde{x})\\
        y_{++}(x;\tilde{x})&=\frac{\Delta x_{++}^2(x;\tilde{x})}{\eta\tilde{\eta}}\\
        \Delta x_{++}^2(x;\tilde{x})&=-(|\eta-\tilde{\eta}|-i\varepsilon)^2+||\vec{x}-\vec{\tilde{x}}||^2,
    \end{split}
\end{equation}
where $Y(X;\tilde{X})=y(x;\tilde{x})$,
$Z(X;\tilde{X})=z(x;\tilde{x})$,  are the functions given in
(\ref{dsfunctions}). $a=a(\eta)$, $\tilde{a}=a(\tilde{\eta})$ and
$\varepsilon>0$ refers to the Feynman (time ordered) pole
prescription. In order to apply the Schwinger-Keldysh (see e.g.
the appendix of \cite{Weinberg:2005vy}) formalism we also define
\begin{equation}
    \begin{split}
        \Delta x_{+-}^2(x;\tilde{x})&=-(\eta-\tilde{\eta}+i\varepsilon)^2+||\vec{x}-\vec{\tilde{x}}||^2\\
        \Delta x_{-+}^2(x;\tilde{x})&=-(\eta-\tilde{\eta}-i\varepsilon)^2+||\vec{x}-\vec{\tilde{x}}||^2\\
        \Delta x_{--}^2(x;\tilde{x})&=-(|\eta-\tilde{\eta}|+i\varepsilon)^2+||\vec{x}-\vec{\tilde{x}}||^2,
    \end{split}
\end{equation}
from which $y_{+-}$, $y_{-+}$ and $y_{--}$ follow immediately. For
the rest of this paper we use
\begin{equation}
    y\equiv y_{++},\qquad\Delta x^2\equiv \Delta x^2_{++}
\end{equation}
unless stated otherwise. When $\varepsilon=0$, the function $y$ is
related to the geodesic length $\ell=\ell(x;\tilde{x})$ between
points $x$ and $\tilde{x}$ as,
\begin{equation}\label{geolength}
 y(x;\tilde{x}) = 4\sin^2\left(\frac12 H\ell(x;\tilde{x})\right)
\,.
\end{equation}
The function $y$ has the following properties: if $y=0$,
$\tilde{x}$ lies on the lightcone of $x$. For $y<0$, $\tilde{x}$
lies in the future or past of $x$ and for $y>0$, $x$ and
$\tilde{x}$ are space-like separated. If $y=4$, $\tilde{x}$ lies
on the light cone of the antipodal point\footnote{The antipodal
point is obtained by the transformation of the coordinates
(\ref{coordinates}) $X_a\rightarrow -X_a$, with $a=(0\ldots4)$} of
$x$ and if $y>4$, $\tilde{x}$ is in the future or past of the
antipodal point.

\subsection{Quasi de Sitter space}\label{sQDS}
Quasi de Sitter space is defined by the requirement that $H$
depends only mildly on time:
\begin{equation}\label{H}
    H= H_0-H_0^2\epsilon
    t+\mathcal{O}(\epsilon^2),\qquad\qquad\epsilon\equiv-\frac{\dot{H}}{H^2}\ll1,\qquad\qquad
    \dot{\epsilon}=\mathcal{O}(\epsilon^2).
\end{equation}
While such a Hubble parameter does not correspond to an exact
solution of Einstein's equations, it is a good approximation in
most inflationary models. For example in scalar field models,
where inflation is achieved by a minimally coupled scalar field
$\phi$, slowly rolling in a potential $V$, $\epsilon$ is just the
slow roll parameter
\begin{equation}
    \epsilon=\frac{1}{2 M_p^2}\Big(\frac{d(V(\phi))/d\phi}{V(\phi)}\Big)^2
\end{equation}
and (\ref{H}) is correct up to order
$\epsilon^2$~\cite{Liddle:1999mq}. The metric of quasi de Sitter
space can be written in conformal coordinates as
\begin{equation}
    g_{\mu\nu}=a^2 \eta_{\mu\nu}
\end{equation}
where the scale factor is given by
\begin{equation}\label{QDS_a1}
    a=-\frac{1}{H_0\eta}\Big(1+(1+\ln(a))\epsilon\Big)+\mathcal{O}(\epsilon^2)
\end{equation}
Since we know from (\ref{dsmetric}) that
$t=\ln(a)/H_0+\mathcal{O}(\epsilon)$, this is equal to
\begin{equation}\label{QDS_a2}
    a=-\frac{1}{\eta H}(1+\epsilon)+\mathcal{O}(\epsilon^2)
\end{equation}
If we take flat spacial slicings, the symmetry group of quasi de
Sitter space is $E(3)$, which corresponds to spatial homogeneity
and isotropy (invariance under translations and rotations).

\section{The scalar propagator}\label{sscal}
The Feynmann propagator
\begin{equation}
    \imath\Delta(x;\tilde{x})\equiv\imath\Delta(x;\tilde{x})_{++}
\end{equation}
of a minimally coupled\footnote{Our results straightforwardly
generalize to a non-minimally coupled scalar by replacing
$m^2\rightarrow m^2+\xi R$, with $R$ the Ricci scalar and $\xi$ a
a coupling parameter.} scalar field with mass $m$ satisfies the
following Klein-Gordon equation:
\begin{equation}\label{Scal_KG}
    \sqrt{-g}(\Box-m^2)\imath\Delta(x;\tilde{x})=\imath\delta^D(x-\tilde{x}),
\end{equation}
where $\Box$ denotes the d'Alembertain
\begin{equation}\label{BOX}
    \Box\equiv \nabla^\alpha\nabla_\alpha
\end{equation}
and $\delta^D$ is the $D$-dimensional Dirac delta. In de Sitter
space, one can solve for $\Delta(x;\tilde{x})$ purely in terms of
the de Sitter invariant distance function $y$ defined in
(\ref{QDS_y})~\cite{Chernikov:1968zm}\cite{Allen:1985ux}.
Replacing $y_{++}$ with $y_{--}$ gives the anti-time ordered
propagator (which has the opposite sign in front of the delta
function), while replacing it with $y_{+-}$ and $y_{-+}$ gives the
two Wightman propagators. The result, known as the
Chernikov-Tagirov propagator\cite{Chernikov:1968zm}, however
becomes singular in the limit when the mass goes to zero. It was
shown in ~\cite{Allen:1985ux}\cite{Allen:1987tz} that it is not
possible to construct a physically meaningful de Sitter invariant
propagator for the massless, minimally coupled scalar field. The
reason is that in such a case the solution to (\ref{Scal_KG}) has
an additional singularity when $\tilde{x}$ is on the lightcone of
the antipodal point of $x$. Such a solution corresponds to the so
called $\alpha$-vacuum. Although the issue is disputed, these
vacua are usually considered
unphysical \cite{Einhorn:2002nu}\cite{Kaloper:2002cs}\cite{Banks:2002nv}~\cite{Danielsson:2002mb}~\cite{Goldstein:2003ut}.\\
In quasi de Sitter we find one can do something similar. It turns
out that after a simple rescaling, we can construct a propagator
purely in terms of the function $y$. Due to $\epsilon$
corrections, this propagator now \emph{has} a consistent massless
limit. The $y$ function we use is the one defined in
(\ref{QDS_y}). One could think however of different definitions
for $y$. For example the most general definition appears to be
through the geodesic length (\ref{geolength}). However the
geodesic length in quasi de Sitter turns out to be a rather
complex object and not very helpful in solving (\ref{Scal_KG}). An
alternative definition of $y$ is
\begin{equation}
    y_{\rm alt}(x;\tilde{x})=a\tilde{a}H\tilde{H}\Delta
    x^2(x;\tilde{x})
\end{equation}
While this definition is equivalent to (\ref{QDS_y}) in de Sitter
space, it differs by a factor $(1+2\epsilon)$ in quasi de Sitter.
We use the definition (\ref{QDS_y}), since in that case the
singularities of the general solution to (\ref{Scal_KG}) are at
$y=0$ and $y=4$ and thus resembles the de Sitter solution most.\\
Our rescaling of the propagator is
\begin{equation}
    \Delta(x;\tilde{x})\rightarrow(a\tilde{a})^{\epsilon w}\Xi(y),
\end{equation}
where $w$ is a constant parameter. We shall determine $w$ by the
requirement that $\Xi$ is a function of $y$ only (strictly
speaking $\Xi$ is not a function, but a distribution). We rewrite
the left hand side of (\ref{Scal_KG}) in terms of $y$-derivatives:
\begin{equation}\label{scalar_prop_eq_temp}
    \begin{split}
    a^{D-2}(a\tilde{a})^{\epsilon w}\Bigg[&\big(-(\partial_0y)^2+(\partial_i
    y)^2\big)\big(\frac{d}{dy}\big)^2\\
    &+\Big[\big(-\partial_0^2y+\partial_i^2y\big)-\big(D-2+2\epsilon w\big)\big(\frac{a'}{a}\big)(\partial_0y)\Big]\frac{d}{dy}-a^2\mu^2\Bigg]\imath\Xi(y)
    \end{split}
\end{equation}
where
\begin{equation}
    \mu^2\equiv
    (D-1)\epsilon wH_0^2+m^2+\mathcal{O}(\epsilon^2)
\end{equation}
and we used
\begin{equation} \label{as}
        \frac{a'}{a^2}=H,\qquad\frac{a''}{a^3}=\dot{H}+2H^2.
\end{equation}

 An important point of this calculation is that, even
if the mass of the field is zero, the effective mass $\mu$ as it
appears in (\ref{scalar_prop_eq_temp}) is nonzero. Therefore we
find the propagator of a massless scalar field in quasi de Sitter
space does not suffer from the presence of $\alpha$-vacua. Indeed
this is expected, since in order to construct the propagator for
such a field in de Sitter space, one exactly breaks the de Sitter
symmetry to the symmetry of quasi de Sitter space. The several
terms in (\ref{scalar_prop_eq_temp}) evaluate to
\begin{equation}
    \begin{split}
    a^{-2}\Big[-(\partial_0y)^2+(\partial_i
    y)^2\Big]=&H^2(1-2\epsilon)y(4-y)+\mathcal{O}(\epsilon^2)\\
    a^{-2}\Big[\big(-\partial_0^2y+\partial_i^2y\big)-\big(D-2+2\epsilon
    w\big)\big(\frac{a'}{a}\big)(\partial_0y)\Big]=&-H^2(1-2\epsilon)\Big(D(y-2)\\
    &+\epsilon(D-2+2w)y\Big)
    \end{split}
\end{equation}
where we made use of (\ref{QDS_a1}) and (\ref{QDS_a2}). The
$\delta$-function is only sourced by the singular term in $\Xi$
for $y\rightarrow 0$. Putting everything together we find for the
nonsingular terms that
\begin{equation}\label{scalar_prop_eq_temp2}
    \Bigg[y(4-y)\big(\frac{d}{dy}\big)^2-\big(D(y-2)+(D-2+2w)y\epsilon\big)\frac{d}{dy}-(1+2\epsilon)\frac{\mu^2}{H^2}\Bigg]\imath\Xi(y)=0.
\end{equation}
In order for $\Xi$ to be a function of $y$ only, the left hand
side of (\ref{scalar_prop_eq_temp2}) needs to be a function of $y$
only. The term $\mu^2/H^2$ can in general never be written as a
function of $y$. However if we assume that our field is light,
such that
\begin{equation}
    m^2=\mathcal{O}(\epsilon),
\end{equation}
$\mu^2/H^2$ is, up to order $\epsilon^2$ corrections, proportional
to $\mu^2/H_0^2$ and thus a constant. The resulting equation is a
hypergeometric equation with a general solution
\begin{equation}\label{generalhyper}
    \begin{split}
        \imath\Xi(y)&=A\,{}_2F_1\big[\frac{D-1}{2}+\frac{D-2+2w}{2}\epsilon+\nu_D,\frac{D-1}{2}+\frac{D-2+2w}{2}\epsilon-\nu_D;\frac{D}{2};\frac{y}{4}\big]\\
        &+B\,{}_2F_1\big[\frac{D-1}{2}+\frac{D-2+2w}{2}\epsilon+\nu_D,\\
        &\qquad\qquad\frac{D-1}{2}+\frac{D-2+2w}{2}\epsilon-\nu_D;\frac{D}{2}+(D-2+2w)\epsilon;1-\frac{y}{4}\big],
    \end{split}
\end{equation}
where
\begin{equation}\label{scalnuorig}
    \nu_D^2=\Big(\frac{D-1}{2}+\frac{D-2+2w}{2}\epsilon\Big)^2-(1+2w)\frac{\mu^2}{H_0^2}.
\end{equation}
The constants $A$ and $B$ can be fixed by looking at the singular
terms. First of all we require that the propagator has no
singularity at $y=4$. Such a singularity would correspond to an
additional delta-source at the antipodal point ($\alpha$-vacuum)
and is considered unphysical. This requirement fixes $A=0$. We fix
$B$ by requiring that the singularity at $y=0$ sources the
$\delta$-function correctly. The $\delta$ function is sourced by
the most singular term, which is given by
\begin{equation}
    \imath \Xi(y)_{\rm
    sing}=B\Big(\frac{y}{4}\Big)^{1-D/2}\frac{\Gamma[D/2+(D-2+2w)\epsilon]\Gamma[D/2-1]}{\Gamma[\frac{D-1}{2}+\frac{D-2+2w}{2}\epsilon+\nu_D]\Gamma[\frac{D-1}{2}+\frac{D-2+2w}{2}\epsilon-\nu_D]}.
\end{equation}
The delta function is sourced by letting $\sqrt{-g}\Box$ act on
this term. We find
\begin{equation}
    \begin{split}
    &\imath\delta^D(x-\tilde{x})=(a\tilde{a})^{\epsilon w}a^{D-2}\partial^2     \imath \Xi(y)_{\rm
    sing}\\
    &=\frac{(a)^{2\epsilon
    w}\Gamma[D/2+(D-2+2w)\epsilon]\Gamma[D/2-1]}{\Gamma[\frac{D-1}{2}+\frac{D-2+2w}{2}\epsilon+\nu_D]\Gamma[\frac{D-1}{2}+\frac{D-2+2w}{2}\epsilon-\nu_D]}B2^{D-2}(-a\eta)^{D-2}\partial^2(\frac{1}{\Delta x^{D-2}})
    \end{split}
\end{equation}
where $\Delta x^2 $ is defined in (\ref{QDS_y}). Depending on the
propagator we calculate, we use one of the following relations
\begin{equation}
    \begin{split}
        \partial^2(\frac{1}{\Delta x_{++}^{D-2}})&=\frac{4\pi^{D/2}}{\Gamma[D/2-1]}\imath\delta^D(x-\tilde{x})\\
        \partial^2(\frac{1}{\Delta x_{--}^{D-2}})&=-\frac{4\pi^{D/2}}{\Gamma[D/2-1]}\imath\delta^D(x-\tilde{x})\\
        \partial^2(\frac{1}{\Delta x_{-+}^{D-2}})&=\partial^2(\frac{1}{\Delta
        x_{+-}^{D-2}})=0.
    \end{split}
\end{equation}
If one uses the appropriate form of (\ref{Scal_KG}) they all give
the following result
\begin{equation}\label{constantB}
    B=\frac{\Gamma[\frac{D-1}{2}+\frac{D-2+2w}{2}\epsilon+\nu_D]\Gamma[\frac{D-1}{2}+\frac{D-2+2w}{2}\epsilon-\nu_D]}{\Gamma[D/2+(D-2+2w)\epsilon]}\frac{H^{D-2}}{(4\pi)^{D/2}(1+\epsilon)^{D-2}a^{2\epsilon
    w}}.
\end{equation}
We can now fix $w$ in order that $B$ is independent of $a$ and
$\eta$. To do this we write
\begin{equation}
    \frac{H^{D-2}}{a^{2w\epsilon}}=H_0^{D-2}(1-\epsilon(D-2+2w)\ln(a))+\mathcal{O}(\epsilon^2)
\end{equation}
and therefore if we choose
\begin{equation}
    w=1-\frac{D}{2}
\end{equation}
$\Xi$ is truly only a function of $y$ up to order $\epsilon^2$
corrections. Remarkably, for this choice of $w$, equations
(\ref{scalar_prop_eq_temp2}) and (\ref{generalhyper}) reduce
exactly to their analogues in de Sitter space, with only the mass
term modified. The final propagator is
\begin{equation}\label{scal_prop}
    \begin{split}
      \imath\Delta(x;\tilde{x})=&\frac{(a\tilde{a})^{-\epsilon(D/2-1)}}{(1+\epsilon)^{D-2}}\frac{H_0^{D-2}}{(4\pi)^{D/2}}\frac{\Gamma[\frac{D-1}{2}+\nu_D]\Gamma[\frac{D-1}{2}-\nu_D]}{\Gamma[\frac{D}{2}]}\\
      &\times{}_2F_1\Big[\frac{D-1}{2}+\nu_D,\frac{D-1}{2}-\nu_D;\frac{D}{2};1-\frac{y}{4}\Big],
      \end{split}
\end{equation}
with
\begin{equation}\label{scal_nu}
    \nu_D^2=\Big(\frac{D-1}{2}\Big)^2-\frac{m^2}{H_0^2}+\frac{\epsilon}{2}(D-1)(D-2),\qquad m^2=\mathcal{O}(\epsilon).
\end{equation}
The $\Delta_{--}$, $\Delta_{+-}$ and $\Delta_{-+}$ propagators can
be obtained by replacing $y_{++}$ in (\ref{scal_prop}) by
$y_{--}$, $y_{+-}$ and $y_{-+}$, respectively. The reason is that
the constant $B$, given in (\ref{constantB}), is the same for all
propagators.
\subsection{The de Sitter limit}
In the limit $\epsilon \rightarrow 0$, the leading order term for
the massive scalar propagator is
\begin{equation}
    \begin{split}
      \lim_{\epsilon\rightarrow
      0}{\imath\Delta(x;\tilde{x})}=&\frac{H_0^{D-2}}{(4\pi)^{D/2}}\frac{\Gamma[\frac{D-1}{2}+\nu_D]\Gamma[\frac{D-1}{2}-\nu_D]}{\Gamma[\frac{D}{2}]}\\
      &\times{}_2F_1\Big[\frac{D-1}{2}+\nu_D,\frac{D-1}{2}-\nu_D;\frac{D}{2};1-\frac{y}{4}\Big]+\mathcal{O}(\epsilon)
      \end{split}
\end{equation}
with
\begin{equation}
    \lim_{\epsilon\rightarrow 0}{\nu_D^2}=\Big(\frac{D-1}{2}\Big)^2-\frac{m^2}{H_0^2}.
\end{equation}
This is -- as expected -- the Chernikov-Tagirov
propagator~\cite{Chernikov:1968zm}. For a massless scalar the
story is slightly more complicated, since the term
\begin{equation}
    \Gamma[\frac{D-1}{2}-\nu_D]
\end{equation}
becomes singular. We split off this singular term by rewriting the
hypergeometric function in  (\ref{scal_prop}) as a series:
\begin{equation}
    \begin{split}
      \Delta(x,\tilde{x})=&\frac{(a\tilde{a})^{-\epsilon(D/2-1)}}{(1+\epsilon)^{D-2}}\frac{H_0^{D-2}}{(4\pi)^{D/2}}\frac{\Gamma[D/2]\Gamma[1-D/2]}{\Gamma[1/2+\nu_D]\Gamma[1/2-\nu_D]}\\
      &\times\Bigg[\sum_{n=1}^\infty\frac{\Gamma[\frac{D-1}{2}+\nu_D+n]\Gamma[\frac{D-1}{2}-\nu_D+n]}{\Gamma[D/2+n]\Gamma[n+1]}\Big(\frac{y}{4}\Big)^n\\
      &\quad-\sum_{n=-1}^\infty\frac{\Gamma[\frac{3}{2}+\nu_D+n]\Gamma[\frac{3}{2}-\nu_D+n]}{\Gamma[3-D/2+n]\Gamma[n+2]}\Big(\frac{y}{4}\Big)^{n-D/2+2}\\
      &\quad+\frac{\Gamma[\frac{D-1}{2}+\nu_D]\Gamma[\frac{D-1}{2}-\nu_D]}{\Gamma[D/2]}\Bigg].
    \end{split}
\end{equation}
Now we can easily expand around $\epsilon=0$ and obtain
\begin{equation}\label{epszero}
    \begin{split}
      \imath\Delta(x;\tilde{x})_{m=0}=&\frac{H_0^{D-2}}{(4\pi)^{D/2}}\Bigg[\frac{-2\Gamma[D-1]}{(D-2)\Gamma[D/2]}\frac{1}{\epsilon}\\
      &+\frac{\Gamma[D-1]}{\Gamma[D/2]}\Big(\ln(a\tilde{a})+2-\gamma_E+\psi(D-1)+\pi\cot\big(\frac{D\pi}{2}\big)\Big)\\
      &+\sum_{n=1}^\infty\frac{\Gamma[n+D-1]}{n\Gamma[D/2+n]}\Big(\frac{y}{4}\Big)^n\\
      &-\sum_{n=-1}^\infty\frac{\Gamma[n+D/2+1]}{(n-D/2+2)\Gamma[n+2]}\Big(\frac{y}{4}\Big)^{n-D/2+2}\Bigg]+\mathcal{O}(\epsilon),
    \end{split}
\end{equation}
where $\gamma_E$ is the Euler-Mascheroni constant and
$\psi[z]\equiv\frac{d}{dz}\ln(\Gamma[z])$ denotes the digamma
function. Apart from the finite constants $2-\gamma_E+\psi(D-1)$
and the $1/\epsilon$ term (which is also a constant in our
approximation), (\ref{epszero}) exactly corresponds to the
massless scalar propagator in~\cite{Onemli:2002hr}. Since a
constant corresponds to a solution of the homogeneous equation
(scalar condensate), it can always be added. This can be seen from
the fact that in the massless case the hypergeometric function
multiplying `$A$' in (\ref{generalhyper}) is to leading order just
1 and thus we are not forced to put $A=0$. The propagator
in~\cite{Onemli:2002hr} is calculated by requiring that the
inevitable breaking of the de Sitter symmetry (to avoid
$\alpha$-vacua) by a massless scalar, is such that it does not
break spacial homogeneity and isotropy (this is known as the
$E(3)$ vacuum). Since those symmetries are the symmetries of quasi
de Sitter space, it is expected that the de Sitter limit of the
quasi de Sitter propagator would correspond up to a constant to
(\ref{epszero}).

\section{The graviton propagator}\label{sgrav}
Various attempts have been made in the past to calculate the
graviton propagator in de Sitter space. In many
works~\cite{Floratos:1987ek}\cite{Allen:1986ta}\cite{Allen:1986tt}\cite{Higuchi:2001uv}
this is done by adding a de Sitter invariant gauge fixing term.
However, it turns out that these Green's functions give a
divergent response to a point mass and moreover they do not solve
the classical invariant equation of
motion\cite{Antoniadis:1986sb}. Based on this Tsamis and Woodard
conclude that there must be some deep, still not completely
understood physical principle that forbids the use of de Sitter
invariant gauge fixing
terms\cite{Woodard:2004ut}\cite{Tsamis:1992xa}\cite{Miao:2005am}.
This problem might be related to the issues concerning the
linearization instability in de Sitter
space\cite{Woodard:2004ut}\cite{Tsamis:1992xa}\cite{Penrose:1964ge}\cite{Bicak:2001fg}\cite{Bicak:2005yt}.
The point is that the linearization instability implies a
constraint on the (classical) field equations. This constraint has
a nonzero response to sources on the whole de Sitter manifold.
However, if one then adds a de Sitter invariant gauge fixing term
and calculates Green's functions, one finds that these Green's
functions only respond to sources in the past light cone. Since
this does not cover the whole manifold, one has changed classical
physics.\footnote{The problem seems to arise as soon as one
imposes an average gauge condition, as when one adds a gauge
fixing term. In Proca theory the de Sitter invariant propagator
has been calculated using an exact gauge condition
\cite{Tsamis:2006gj}. This propagator does give the correct
response.}
\\
This problem is circumvented by working only on half of the de
Sitter manifold and by using a gauge fixing term that breaks the
de Sitter symmetry\cite{Tsamis:1992xa}. If one calculates the
graviton propagator in this way, one finds that the solution can
be written in terms of three scalar propagators, multiplying some
tensor structure \cite{Tsamis:2005je}\cite{Tsamis:1992xa}. Since
one of these propagators corresponds to a massless minimally
coupled scalar, the graviton propagator inevitably breaks the de
Sitter symmetry.\\
In this section we calculate the graviton propagator in quasi de
Sitter space. Although slightly more complicated, the result is
similar as in de Sitter space. Two differences are that our gauge
fixing term, the analogue of the one used in
\cite{Tsamis:2005je}\cite{Tsamis:1992xa}, does not break the quasi
de Sitter symmetry. Moreover, as we have shown above, a massless
minimally coupled scalar poses no problems in quasi de Sitter space.\\

\subsection{The graviton action}
We start our calculation with the Einstein-Hilbert action with a
cosmological constant $\Lambda$, coupled to a scalar field $\phi$.
\begin{equation}\label{action}
    S_{\rm
    EH}=\frac{1}{\kappa}\int\sqrt{-\hat{g}}\Big(\hat{R}-(D-2)\Lambda\Big)+S_M,
\end{equation}
where
\begin{equation}\label{Sm}
    \begin{split}
        \kappa&=16\pi G_N\\
        S_M&=\int\sqrt{-g}\Big(-\frac{1}{2}\partial_\alpha\phi\partial_\beta\phi
        g^{\alpha\beta}-V(\phi)\Big)
    \end{split}
\end{equation}
and the hats indicate `background+perturbation'.
 We have added the matter action $S_M$ for
consistency, since without it, the quasi de Sitter metric is not a
solution to Einstein's equations. Notice that the addition of
matter induces a mixing between the graviton and the scalar field.
Taking this mixing into account is crucial for the calculation of
scalar cosmological perturbations~\cite{Mukhanov:2005sc}. The four
dimensional propagators, with mixing, are calculated
in~\cite{Iliopoulos:1998wq}\cite{Abramo:2001dc} in power law
expansion. A study of the mixing between the gravitational and the
matter section in $D$ dimensions will be done
elsewhere~\cite{inprogress}. Quasi de Sitter is a good
approximation if the scalar field $\phi$ is slowly rolling in the
potential $V$ \cite{Liddle:1999mq}. Using the geometric quantities
from appendix \ref{appgeo}, the Friedmann and fluid equation
corresponding to (\ref{action}) are found to be
\begin{equation}\label{Fried}
    \begin{split}
        &H^2=\frac{1}{D-1}\Lambda+\frac{\kappa}{(D-1)(D-2)}(\frac{1}{2}\dot{\phi}^2+V(\phi))\\
        &\ddot{\phi}+(D-1)H\dot{\phi}=-\frac{d V(\phi)}{d\phi},
    \end{split}
\end{equation}
where a dot denotes a time derivative. By choosing the appropriate
potential, (\ref{Fried}) can describe any time evolution of the
scale factor. The action (\ref{action}) therefore describes -- as
far as the geometry is concerned -- a completely general FLRW
universe. To calculate the graviton action we decompose our metric
in a background $g$ plus a perturbation
$h$~\cite{Christensen:1979iy}:
\begin{equation}\label{metricdecomp}
\begin{split}
    &\hat{g}_{\mu\nu}=g_{\mu\nu}+h_{\mu\nu}\\
    &\hat{g}^{\mu\nu}=g^{\mu\nu}+\delta g^{\mu\nu}
\end{split}
\end{equation}
If we require that
$\hat{g}^{\mu\alpha}\hat{g}_{\mu\beta}=\delta^\alpha_\beta$ and we
raise and lower indices on the perturbation with the background
metric: $h^\mu{}_\alpha=g^{\mu\nu}h_{\nu\alpha}$, we find that
\begin{equation}
    \delta g^{\mu\nu}=-h^{\mu\nu}+h^\mu{}_\alpha
    h^{\alpha\nu}+\mathcal{O}(h^3).
\end{equation}
We expand the action in powers of the perturbation around the
background as
\begin{equation}
    S=S_0+S_1+ \frac{1}{2}S_2 +...,
\end{equation}
where it should be noted that, since $\delta g^{\mu\nu}$ is
quadratic in $h$, $S_1$ contains a part quadratic in $h$. For the
first two terms we get
\begin{equation}
    \begin{split}
        S_0&=\frac{1}{\kappa}\int\sqrt{-g}\Big(R-(D-2)\Lambda\Big)+S_M\\
        S_1&=-\frac{1}{\kappa}\int\sqrt{-g}\Big((EE)_{\mu\nu}h^{\mu\nu}-(EE)_{\mu\nu} h^{\mu}{}_\alpha
            h^{\alpha\nu}\Big)
    \end{split}
\end{equation}
where
\begin{equation}\label{tmn}
    \begin{split}
        T_{\mu\nu}&\equiv-\frac{2}{\sqrt{-g}}\frac{\delta
        S_M}{\delta g^{\mu\nu}}=(\partial_\mu\phi)(\partial_\nu\phi)-g_{\mu\nu}\Big(\frac{1}{2}(\partial_\alpha\phi)(\partial_\beta\phi) g^{\alpha\beta}+V(\phi)\Big)\\
        (EE)_{\mu\nu}&\equiv R_{\mu\nu}-\frac{1}{2}g_{\mu\nu}R+\frac{D-2}{2}\Lambda
        g_{\mu\nu}-\frac{\kappa}{2} T_{\mu\nu}.
    \end{split}
\end{equation}
If we drop terms cubic in $h$, $S_2$ is given by
\begin{equation}
    \begin{split}
        S_2
        =-\frac{1}{\kappa}\int\sqrt{-g}\Bigg[&\frac{1}{2} h
        h^{\mu\nu}(EE)_{\mu\nu}+\tilde{h}^{\mu\nu}(\delta
        R_{\mu\nu})-\frac{1}{2} R h_{\mu\nu}
        h^{\mu\nu}+\frac{1}{2} h h^{\mu\nu} R_{\mu\nu}\\
        &+\frac{D-2}{2}\Lambda h_{\mu\nu}
        h^{\mu\nu}-\frac{\kappa}{2}h^{\mu\nu}(\delta
        T_{\mu\nu})\Bigg],
    \end{split}
\end{equation}
where we have defined the trace reversed graviton:
\begin{equation}\label{tracereversed}
    \tilde{h}_{\mu\nu}\equiv h_{\mu\nu}-\frac{1}{2}g_{\mu\nu}
    h,\qquad     h\equiv g^{\mu\nu}h_{\mu\nu}.
\end{equation}
Using the following identities~\cite{Carroll:1997ar}
\begin{equation}
    \delta
    R_{\alpha\beta}=\frac{1}{2}g^{\lambda\rho}\Big(\nabla_\lambda\nabla_\alpha
    h_{\rho\beta}+\nabla_\lambda\nabla_\beta
    h_{\rho\alpha}-\nabla_\lambda\nabla_\rho
    h_{\alpha\beta}-\nabla_\alpha\nabla_\beta h_{\lambda\rho}\Big)
\end{equation}
\begin{equation}
    \tilde{h}^{\alpha\beta}g^{\lambda\rho}\nabla_\lambda\nabla_\alpha
    h_{\rho\beta}=\tilde{h}^{\alpha\beta}g^{\lambda\rho}\nabla_\alpha\nabla_\lambda
    h_{\rho\beta}+R_{\sigma\alpha}h^\sigma{}_\beta\tilde{h}^{\alpha\beta}-R_{\beta\sigma\alpha\lambda}h^{\sigma\lambda}\tilde{h}^{\alpha\beta}
\end{equation}
and removing boundary terms, we find
\begin{equation}
    \begin{split}
        \frac{S_2}{2}=&\frac{1}{\kappa}\int\sqrt{-g}\Bigg[\frac{1}{4}\tilde{h}^{\mu\nu}\Big(\Box
        g_{\alpha\mu}g_{\beta\nu}+2
        R_{\alpha\mu\beta\nu}\Big)h^{\alpha\beta}+\frac{1}{2}(\nabla_\mu\tilde{h}^{\mu\alpha})(\nabla^\nu\tilde{h}_{\nu\alpha})\\
        &+\frac{\kappa}{4}h^{\mu\nu}(\delta T_{\mu\nu})-\frac{\kappa}{4} T_{\mu\nu} h^{\alpha\nu} h_\alpha{}^\mu
        -\frac{1}{2}\Big(h^{\alpha\nu}h_\alpha{}^\mu+\frac{1}{2}h
        h^{\mu\nu}\Big)(EE)_{\mu\nu}\Bigg],
    \end{split}
\end{equation}
where the d'Alembertian is given in (\ref{BOX}). Adding the part
quadratic in $h$ from $S_1$ the total quadratic part of the action
becomes:
\begin{equation}\label{generalquad}
    \begin{split}
        S^{(2)}=\frac{1}{\kappa}\int\sqrt{-g}\Bigg[&\frac{1}{4}\tilde{h}^{\mu\nu}\Big(\Box
        g_{\alpha\mu}g_{\beta\nu}+2
        R_{\alpha\mu\beta\nu}\Big)h^{\alpha\beta}+\frac{1}{2}(\nabla_\mu\tilde{h}^{\mu\alpha})(\nabla^\nu\tilde{h}_{\nu\alpha})\\
        &+\frac{\kappa}{4}\Big(h^{\mu\nu}(\delta T_{\mu\nu})-T_{\mu\nu} h^{\alpha\nu}
        h_\alpha{}^\mu\Big)+\frac{1}{2}\tilde{h}^{\rho\nu}h_\rho{}^\mu(EE)_{\mu\nu}\Bigg]
    \end{split}
\end{equation}

In the following we drop the $(EE)_{\mu\nu}$ terms from
(\ref{generalquad}), since they are zero on-shell. The
contribution to the graviton action from the matter action can be
found from (\ref{tmn}) and
\begin{equation}
    \delta T_{\mu\nu}=-h_{\mu\nu}\Big(\frac{1}{2}(\partial_\alpha\phi)(\partial_\beta\phi)
    g^{\alpha\beta}+V(\phi)\Big)+\frac{1}{2}g_{\mu\nu}(\partial_\alpha\phi)(\partial_\beta\phi)
    h^{\alpha\beta}
\end{equation}
to give
\begin{equation}
    \frac{\kappa}{4}\Big(h^{\mu\nu}(\delta T_{\mu\nu})-T_{\mu\nu} h^{\alpha\nu}
        h_\alpha{}^\mu\Big)=\frac{\kappa}{8}(\phi')^2\Big(h
        h^{00}-2h^{0\lambda} h^0{}_\lambda\Big)
\end{equation}
From (\ref{Fried}) we can derive that
\begin{equation}
    \kappa(\phi')^2=-2(D-2)\dot{H}a^2,
\end{equation}
where a prime denotes a derivative with respect to conformal time.
Using this, the graviton action becomes
\begin{equation}
    \begin{split}
        S^{(2)}_g=\frac{1}{\kappa}\int\sqrt{-g}\Bigg[&\frac{1}{4}\tilde{h}^{\mu\nu}\Big(\Box
        g_{\alpha\mu}g_{\beta\nu}+2
        R_{\alpha\mu\beta\nu}\Big)h^{\alpha\beta}+\frac{1}{2}(\nabla_\mu\tilde{h}^{\mu\alpha})(\nabla^\nu\tilde{h}_{\nu\alpha})\\
        &+\frac{D-2}{4}\dot{H}\Big(h h^0{}_0-2h^{\lambda 0}h_{\lambda 0}\Big)\Bigg].
    \end{split}
\end{equation}
In any FLRW metric we can choose coordinates where the metric is
conformally flat,
\begin{equation}
    g_{\mu\nu}=a^2 \eta_{\mu\nu}.
\end{equation}
A lengthy calculation with several partial integrations and using
the quantities given in Appendix \ref{appgeo} gives apart from
boundary terms the following result~\cite{Tsamis:1992xa}:
\begin{equation}
    \begin{split}
    S^{(2)}=a^{D+4}\Bigg[&\frac{1}{4}\psi^{\mu\nu}(\partial^\alpha\partial_\alpha-(D-2)\Big(\frac{a'}{a^3}\Big)\partial_0)\psi_{\mu\nu}\\
    &-\frac{1}{8}\psi\Big(\partial^\alpha\partial_\alpha-(D-2)\Big(\frac{a'}{a^3}\Big)\partial_0+4\Big(\frac{a'}{a^2}\Big)^2\Big)\psi\\
    &-\frac{D-2}{2}\Big(\Big(\frac{a'}{a^2}\Big)^2+\dot{H}\Big)\psi^{0\nu}\psi_{0\nu}+\frac{D-2}{4}\Big(2\Big(\frac{a'}{a^3}\Big)^2-\frac{a''}{a^2}+\dot{H}\Big)\psi^0{}_0\psi\\
    &+\frac{1}{2}(\nabla_\mu\tilde{\psi}^{\mu\alpha})(\nabla_\nu\tilde{\psi}^\nu{}_\alpha)\Bigg],
    \end{split}
\end{equation}
where
\begin{equation}
    h_{\mu\nu}=\sqrt{\kappa}a^2\psi_{\mu\nu},\qquad\tilde{\psi}_{\mu\nu}=\psi_{\mu\nu}-\frac{1}{2}g_{\mu\nu}\psi,\qquad\psi=g^{\mu\nu}\psi_{\mu\nu}.
\end{equation}
We follow~\cite{Tsamis:1992xa} and add the following gauge fixing
term:
\begin{equation}\label{gaugefix}
    -\frac{1}{2}a^{D+4}(\nabla_\mu\tilde{\psi}^{\mu\alpha})(\nabla_\nu\tilde{\psi}^\nu{}_\alpha).
\end{equation}
While this gauge fixing term breaks the de Sitter symmetry, it
does respect the quasi de Sitter symmetry. Removing another total
derivative and, using (\ref{as}), we obtain the general form of
the gauge fixed graviton action in $FLRW$ space-time
\begin{equation} \label{grav_action}
        S_{GF}^{(2)}=\int
        a^{D}\eta^{\alpha\mu}\eta^{\beta\nu}\psi_{\alpha\beta}\Bigg(\Box_S\Big(\frac{1}{4}\delta_\mu^\rho\delta_\nu^\sigma-\frac{1}{8}g_{\mu\nu}g^{\rho\sigma}\Big)-\frac{D-2}{2}(H^2+\dot{H})\delta^0_\mu\delta^\rho_\nu\delta^\sigma_0\Bigg)\psi_{\rho\sigma},
\end{equation}
where
\begin{equation}\label{box2}
    \Box_S=a^{-D}\partial_\alpha
    a^{D-2}\eta^{\alpha\beta}\partial_\beta
\end{equation}
is the d'Alembertian as it acts on a scalar field.
\subsubsection{The Ghost lagrangian}
The ghost Lagrangian corresponding to our gauge fixing
(\ref{gaugefix}) is given by
\begin{equation}\label{ghostlag}
    \mathcal{L}_{\rm ghost}=-a^{D}\bar{V}^\mu\delta F_\mu
\end{equation}
where $\bar{V}$ is the anti-ghost field and
\begin{equation}
    \begin{split}
        F_\mu=&\nabla_\alpha\tilde{\psi}^\alpha{}_\nu\\
        =&\partial_\alpha\psi^\alpha{}_\nu-\frac{1}{2}\partial_\nu\psi+\Big(\frac{a'}{a}\Big)\Big(D\psi^0{}_\nu-\delta^0_\nu\psi\Big).
    \end{split}
\end{equation}
To calculate $\delta F_\mu$ we consider an infinitesimal
coordinate transformation (indicated by $\wr$)
\begin{equation}
    y^{\wr\mu}=y^\mu+\kappa V^\mu.
\end{equation}
Under this transformation the metric $\hat{g}$ (see
(\ref{metricdecomp})) transforms as
\begin{equation}
        \hat{g}^\wr_{\mu\nu}(y)=\hat{g}_{\mu\nu}(y)-\kappa\Big(\hat{g}_{\alpha\nu}(y)\partial_\mu
    V^\alpha+\hat{g}_{\beta\mu}(y)\partial_\nu
    V^\beta+V^\rho \partial_\rho\hat{g}_{\mu\nu}(y)
    \Big)+\mathcal{O}(V^2)
\end{equation}
and thus we find for the variation of the rescaled graviton field
\begin{equation}
    \begin{split}
\delta \psi_{\mu\nu}=&-a^{-2}\Big(g_{\alpha\nu}\partial_\mu
V^\alpha+g_{\alpha\mu}\partial_\nu
V^\alpha+2\Big(\frac{a'}{a}\Big)g_{\mu\nu} V^0\Big)\\
&-\kappa\Big(\psi_{\alpha\nu}\partial_\mu
V^\alpha+\psi_{\alpha\mu}\partial_\nu
V^\alpha+2\Big(\frac{a'}{a}\Big)\psi_{\mu\nu}
V^0+V^\rho\partial_\rho\psi_{\mu\nu}\Big).
\end{split}
\end{equation}
Since we let $V^\mu$ be our anti-commuting ghost field, the second
line will be cubic in the fields, when combined with
(\ref{ghostlag}), while for the purpose of this paper we are only
interested in quadratic contributions. So dropping the second
line, we obtain
\begin{equation}
    \begin{split}
    \delta F_\mu=a^{-2}\Big(&-g_{\alpha\mu}\partial^\nu\partial_\nu
    V^\alpha+(D-2)g_{\alpha\mu}\Big(\frac{a'}{a^3}\Big)\partial_0
    V^\alpha-(D-2)\Big(\frac{a''}{a^3}\Big)g_{0\nu}V^0\\
    &+3(D-2)\Big(\frac{a'}{a^2}\Big)^2g_{0\nu}V^0\Big).
    \end{split}
\end{equation}
Now using (\ref{as}) and (\ref{box2}) we find from
(\ref{ghostlag})
\begin{equation}\label{ghostlagfinal}
    \mathcal{L}_{\rm ghost}=a^D\eta_{\alpha\beta}
    \bar{V}^\beta\Big(\delta^\alpha_\mu\Box_S-(D-2)(H^2-\dot{H})\delta_\mu^0\delta_0^\alpha\Big)V^\mu.
\end{equation}

\subsection{Calculating the propagator}
It is a well known fact that the propagator should be invariant
under the interchange of both legs
\begin{equation}
    \imath\Delta(x;\tilde{x})=\imath\Delta(\tilde{x};x).
\end{equation}
Looking at (\ref{grav_action}) we therefore see that, since
$\Box_S$ does \emph{not} commute with $g_{\mu\nu}$, it acts on the
other leg as
\begin{equation}\label{things}
    \Box_S(\eta^{\alpha\mu}\eta^{\beta\nu}\psi_{\alpha\beta})=\Box_S(a^4\psi^{\mu\nu})
\end{equation}
Thus from  (\ref{grav_action}) we find that the gauge fixed
kinetic operator is
\begin{equation}\label{grav_kinop}
    \mathcal{D}_{\mu\nu}{}^{\rho\sigma}=a^{D}\Bigg[\Box_S\Big(\frac{1}{2}\delta_\mu^{(\rho}\delta_\nu^{\sigma)}-\frac{1}{4}\eta_{\mu\nu}\eta^{\rho\sigma}\Big)-(D-2)(H^2+\dot{H})\delta^0_{(\mu}\delta^{(\rho}_{\nu)}\delta^{\sigma)}_0\Bigg],
\end{equation}
where, because of (\ref{things}) indices are effectively raised
and lowered with $\eta^{\alpha\beta}$. The Feynmann propagator is
defined by
\begin{equation} \label{grav_greenfct}
    \mathcal{D}_{\mu\nu}{}^{\rho\sigma}[\imath_{\rho\sigma}\Delta^{\alpha\beta}](x;\tilde{x})=\imath\delta_\mu^{(\alpha}\delta_\nu^{\beta)}\delta^D(x-\tilde{x}).
\end{equation}
Also indices of the propagator should be raised and lowered with
the Minkowski metric $\eta_{\mu\nu}$. We solve this equation by
making the following ansatz,
\begin{equation}\label{grav_prop_ansatz}
    \begin{split}
    \imath[_{\rho\sigma}\Delta^{\alpha\beta}](x,\tilde{x})=&a(x,\tilde{x})\delta_\rho^{(\alpha}\delta_\sigma^{\beta)}+b(x,\tilde{x})\delta^0_{(\rho}\delta_{\sigma)}^{(\alpha}\delta^{\beta)}_0+c(x,\tilde{x})\eta_{\rho\sigma}\eta^{\alpha\beta}\\
    &+d(x,\tilde{x})\Big(\delta^0_\rho \delta_\sigma^0
    \eta^{\alpha\beta}+\delta^\alpha_0 \delta^\beta_0
    \eta_{\rho\sigma}\Big)+e(x,\tilde{x})\delta_\rho^0\delta_\sigma^0\delta^\alpha_0\delta^\beta_0.
    \end{split}
\end{equation}
This is the most general form that
$[_{\rho\sigma}\Delta^{\alpha\beta}]$ can take, consistent with
the symmetries~\cite{Tsamis:1992xa}. We insert this ansatz in
(\ref{grav_greenfct}) and, combining terms that multiply the same
tensor structure, we obtain 6 differential equations (in the
following $a,b,c,d$ and $e$ are functions of $x$ and $\tilde{x}$)
\begin{eqnarray}\label{grav_diffeq}
    \delta^{(\alpha}_\mu\delta^{\beta)}_\nu\quad:&\quad\frac{1}{2}\Box_S (a)=\imath
    a^{-D}\delta^D(x-x')\\\label{1}
    \delta^0_{(\mu}\delta_{\nu)}^{(\alpha}\delta^{\beta)}_0\quad:&\quad\frac{1}{2}\Box_S(b)-(D-2)(H^2+\dot{H})(a+\frac{1}{2}b)=0\\\label{2}
    \eta_{\mu\nu}\eta^{\alpha\beta}\quad:&\quad\Box_S\Big(\frac{2-D}{4}c-\frac{1}{4}a+\frac{1}{4}d\Big)=0\\\label{3}
    \delta^0_\mu \eta_{\nu
    0}\eta^{\alpha\beta}\quad:&\quad\frac{1}{2}\Box_S(d)+(D-2)(H^2+\dot{H})(c-d)=0\\\label{4}
    \eta_{\mu\nu}\delta^\alpha_0 \eta^{\beta
    0}\quad:&\quad\Box_S\Big(\frac{2-D}{4}d+\frac{1}{4}b+\frac{1}{4}e\Big)=0\\\label{5}
    \delta_\mu^0\delta_\nu^0\delta^\alpha_0\delta^\beta_0\quad:&\quad\frac{1}{2}\Box_S(e)-(D-2)H^2(\frac{1}{2}b-d+e)=0\label{6}.
\end{eqnarray}
To solve this system first of all note that (\ref{2}) and
(\ref{4}) just mean that
\begin{equation}
    \begin{split}
        d&=-(D-2)c-a\\
        e&=-(D-2)d-b=(D-2)^2x+(D-2)a-b
    \end{split}
\end{equation}
So the only independent functions are $a$, $b$ and $c$. We
redefine them as
\begin{equation}\label{grav_abcfunctions}
    \begin{split}
    a&=2\Delta_0\\
    b&=4(\Delta_1-\Delta_0)\\
    c&=\frac{2}{(D-3)(D-2)}(\Delta_2-(D-2)\Delta_0),
    \end{split}
\end{equation}
where $\Delta_{0,1,2}$ are functions of $x$ and $\tilde{x}$.  We
find that the system (\ref{grav_diffeq}...\ref{6}) is solved by
the three functions $\Delta_n$ obeying\cite{Janssen:2007yu}
\begin{equation}\label{scalarprops}
    \Big(\frac{\Box_S}{H^2}-n(D-n-1)(1-\epsilon)\Big)\Delta_n=\frac{\imath\delta^D(x-x')}{a^{D}H^2},
\end{equation}
where $\epsilon$ is defined in (\ref{H}).  Equation
(\ref{scalarprops}) is just the scalar propagator equation
(\ref{Scal_KG}). So, just like in de Sitter
space~\cite{Tsamis:1992xa}, we find that we can write the graviton
propagator in \emph{any} FLRW spacetime in terms of three scalar
propagators, with different couplings to the background curvature.
Inserting (\ref{grav_abcfunctions}) in the ansatz
(\ref{grav_prop_ansatz}), we obtain the graviton propagator in any
FLRW background
\begin{equation}\label{gravprop}
   \begin{split}
    [_{\rho\sigma}\Delta^{\alpha\beta}]=&\Big(2\bar{\delta}^{(\alpha}_\rho\bar{\delta}^{\beta)}_\sigma-\frac{2}{D-3}\bar{\eta}_{\rho\sigma}\bar{\eta}^{\alpha\beta}\Big)\Delta_0+4\delta_{(\rho}^0\bar{\delta}_{\sigma)}^{(\alpha}\delta^{\beta)}_0\Delta_1\\
    &+\Bigg[\frac{2}{(D-2)(D-3)}(\eta_{\rho\sigma}+(D-2)\delta_\sigma^0 \delta_\rho^0)(\eta^{\alpha\beta}+(D-2)\delta^\beta_0\delta^\alpha_0)\Bigg]\Delta_2,
    \end{split}
\end{equation}
where
\begin{equation}
    \bar{\eta}_{\mu\nu}=\eta_{\mu\nu}+\delta_\mu^0 \delta_\nu ^0.
\end{equation}
How to solve (\ref{scalarprops}) in a general FLRW space is not
known\footnote{It is however not hard to solve them if the
equation of state parameter $w\equiv p_\phi/\rho_\phi$ is a
constant, because in those cases, even though $\epsilon$ does not
have to be small, it is constant.}, but from section \ref{sscal}
we know how to solve these type of equations in quasi de Sitter
space. In that case the functions $\Delta_n$ are given by
(\ref{scal_prop}) with
\begin{equation}\label{gravnu}
    \nu_{(D\rm ,
    n)}^2=\frac{(D-2n-1)^2}{4}+\epsilon\Big(\frac{1}{2}(D-1)(D-2)-n(D-n-1)\Big).
\end{equation}
Notice that in calculating (\ref{gravnu}) we did not use
(\ref{scal_nu}), since the requirement $m^2=\mathcal{O}(\epsilon)$
is not met. However, there is in this case no problem in using
(\ref{scalnuorig}) to calculate $\nu_D$, since for the propagators
(\ref{gravprop}) $\mu^2/H^2$ is a constant. The graviton
propagator (\ref{gravprop}) is the main result of this paper.
\subsubsection{Comparison to mode function analysis}
It is an interesting exercise to see how our results compare to
the standard treatment of gravitons in terms of mode functions.
For this we only consider the `physical' graviton $\psi_{ij}$,
defined by requiring that
\begin{equation}
    \psi_{0\mu}=0,\qquad\psi=0.
\end{equation}
We see that in this case the action (\ref{grav_action})leads to
the following equation of motion
\begin{equation}
    \Box_S
    \psi_{ij}=a^{-2}\Big(\partial^2-(D-2)\frac{a'}{a}\partial_0\Big)\psi_{ij}=0
\end{equation}
We rescale $\psi$ to the conformally coupled field $\psi^c$
defined as
\begin{equation}
    \psi_{ij}\equiv a^{-(D-2)/2}\psi_{ij}^c.
\end{equation}
The equation of motion for $\psi^c$ is in Fourier space
\begin{equation}
    \Bigg[-\partial_0^2-k^2+\frac{(D-2)(D-4)}{4}\Big(\frac{a'}{a}\Big)^2+\frac{D-2}{2}\frac{a''}{a}\Bigg]\psi^c_{ij}(k)=0.
\end{equation}
Using (\ref{QDS_a1}) we find that
\begin{equation}
    \Big(\frac{a'}{a}\Big)^2=\frac{1+2\epsilon}{\eta^2},\qquad\frac{a''}{a}=\frac{2+3\epsilon}{\eta^2},
\end{equation}
where $\eta$ is conformal time. The equation of motion reduces
\cite{Mukhanov:2005sc}\cite{Mukhanov:1990me}\cite{Grishchuk:1974ny}
to
\begin{equation}
    \Bigg[\partial_0^2+k^2-\Big(\frac{D(D-2)}{4}+\frac{(D-2)(D-1)}{2}\epsilon\Big)\frac{1}{\eta^2}\Big]\psi^c_{ij}(k)=0.
\end{equation}
This equation can be solved and the result is
\begin{equation}\label{modesol}
    \psi^c_{ij}(k)=c_1 \sqrt{\eta} H^{(1)}_\nu(k\eta)+c_2 \sqrt{\eta}H^{(2)}_\nu(k\eta),
\end{equation}
where $H^{(1,2)}$ are Hankel functions of the first and second
kind respectively and $c_{1,2}$ are constants. The index $\nu$ is
given by
\begin{equation}
    \nu^2=\frac{(D-1)^2}{4}+\epsilon\frac{(D-1)(D-2)}{2}.
\end{equation}
Comparing this result with (\ref{gravnu}), we see that $\nu$ is
exactly equal to $\nu_{(D,0)}$. The reason that we recover the
$n=0$ index, can be seen from (\ref{gravprop}) by the fact that
only $\Delta_0$ acts on the physical graviton $\psi_{ij}$. This
indicates that our propagator correctly reproduces the standard
graviton two-point correlations, obtained from the mode
functions.\\ From this calculation, we can also find a relation
between $\nu$ and $n_g$, the spectral index of graviton
perturbations
\begin{equation}
    n_g=\frac{d\ln{\mathcal{P}_g}}{d\ln{k}}.
\end{equation}
The graviton power-spectrum $\mathcal{P}_g$ is, apart from some
normalization
\begin{equation}
    \mathcal{P}_g\propto k^{D-1}|\psi^c_{ij}(k)|^2.
\end{equation}
We can calculate the power-spectrum in the infrared by expanding
(\ref{modesol}) for small $k$. The result is
\begin{equation}
    \psi_{ij}^c(k)\propto\eta^{1/2}(\eta
    k)^{-\nu}+\mathcal{O}(k^{-\nu+2})
\end{equation}
and therefore we find that
\begin{equation}
    n_g=(D-1)-2\nu=-(D-2)\epsilon+\mathcal{O}(\epsilon^2),
\end{equation}
which in D=4 indeed corresponds to the well known result that
$n_g=-2\epsilon$ \cite{Liddle:1999mq}\cite{Mukhanov:2005sc}.

\subsubsection{Ghost propagator}
By similar arguments as in the previous section, we raise and
lower indices on the ghost field with $\eta^{\mu\nu}$. It follows
from (\ref{ghostlagfinal}) that the kinetic operator is given by
\begin{equation}
    \mathcal{D}_\mu{}^\alpha=a^D\Big(\delta^\alpha_\mu\Box_S-(D-2)(H^2-\dot{H})\delta_\mu^0\delta_0^\alpha\Big).
\end{equation}
The propagator is defined by
\begin{equation}
    \mathcal{D}_\mu{}^\alpha\imath[{}_\alpha\Delta^\rho](x;\tilde{x})=\imath
    \delta^\rho_\mu\delta^D(x-\tilde{x})
\end{equation}
and we find that
\begin{equation}
    \imath[_\alpha\Delta^\rho](x;\tilde{x})=\imath\bar{\delta}_\alpha^\rho\hat{\Delta}_0(x;\tilde{x})+\imath\delta_\alpha^0\delta^\rho_0\hat{\Delta}_1(x;\tilde{x}),
\end{equation}
where the $\hat{\Delta}_n$ propagators satisfy
\begin{equation}\label{ghostprops}
    \Big(\frac{\Box_S}{H^2}-n(D-n-1)(1+\epsilon)\Big)\hat{\Delta}_n=\frac{\imath\delta^D(x-x')}{a^{D}H^2}.
\end{equation}
Just as (\ref{scalarprops}), the functions $\hat{\Delta}$ can be
computed as in section \ref{sscal}. The solution is given by
(\ref{scal_prop}) with
\begin{equation}
    \hat{\nu}_{(D\rm ,
    n)}^2=\frac{(D-2n-1)^2}{4}+\epsilon\Big(\frac{1}{2}(D-1)(D-2)-3n(D-n-1)\Big).
\end{equation}

\section{Conclusion}\label{scon}
While de Sitter space is an excellent framework to do calculations
of quantum effects during inflation, it has two drawbacks. The
first is that it is never realized in nature and the second is
that it is not dynamical in the sense that the Hubble parameter is
constant. To solve these problems one needs to do calculations in
quasi de Sitter space, where first order correction in the slow
roll parameter $\epsilon\equiv-\dot{H}/H^2$ to the background
geometry are taken into account. Not only is quasi de Sitter space
a more realistic approximation of the inflationary universe, it is
also a dynamical space. Indeed, since $\dot{H}$ is nonzero, the
Hubble parameter is allowed to change. Especially if one wants to
consider back-reaction effects or calculate quantum corrections to
tensor cosmological perturbations, taking these dynamics into
account is crucial for self-consistency of the calculation.\\
In this paper the propagator for the graviton in quasi de Sitter
space has been constructed in arbitrary dimension. In order to
achieve this, first the propagator for a scalar field has been
obtained. After a simple rescaling by a factor
$(a\tilde{a})^{(1-D/2)\epsilon}$, the scalar propagator in quasi
de Sitter space has an identical structure as the corresponding
scalar propagator in de Sitter space. The only difference is an
order $\epsilon$ correction to the effective mass. Due to this
correction, the massless limit is regular, where it is singular in
de Sitter space. \\
Just as in de Sitter space, it is found that the graviton
propagator in any FLRW space can be written as the sum of three
scalar propagators with different amounts of coupling to the
background geometry multiplying three different tensorial
structures. If the background is quasi de Sitter, we are able to
explicitly construct these scalar propagators. The graviton
propagator in quasi de Sitter space allows for self-consistent
calculations of graviton back-reaction and for quantum studies of
tensor cosmological perturbations.

\section{Acknowledgements}
We would like to thank Richard Woodard for illuminating
discussions and comments.

\appendix
\section{Geometric quantities}\label{appgeo}
In this appendix we summarize some of the geometric quantities
used in the text. The results in this appendix are valid in any
FLRW spacetime. In this paper we work (mostly) with conformally
flat space-times where the metric is given by
\begin{equation}
    g_{\mu\nu}=a^2 \eta_{\mu\nu}\qquad\qquad
    \eta_{\mu\nu}=\text{diag}(-1,1,1,1)
\end{equation}
We define the covariant derivative as\cite{Carroll:1997ar}
\begin{equation}
    \nabla_\mu V^\alpha_{}\beta=\partial_\mu
    V^\alpha{}_\beta+\Gamma^\alpha_{\mu\lambda}
    V^\lambda{}_\beta-\Gamma^\lambda_{\mu\beta} V^\alpha{}_\lambda
\end{equation}
where the connection coefficients are given by
\begin{equation}\label{connection}
    \Gamma^\alpha_{\mu\nu}=\frac{a'}{a}\Big(\delta^\alpha_\mu\delta^0_\nu+\delta^\alpha_\nu\delta^0_\mu+\delta^\alpha_0\eta_{\mu\nu}\Big),
\end{equation}
and a prime indicates a dervative with respect to conformal time.
The Riemann tensor is defined as
\begin{equation}
    R^\alpha{}_{\mu\beta\nu}=\partial_\beta
    \Gamma^\alpha_{\mu\nu}-\partial_\nu\Gamma^\alpha_{\beta\mu}+\Gamma^\alpha_{\beta\lambda}\Gamma^\lambda_{\mu\nu}-\Gamma^\alpha_{\nu\lambda}\Gamma^\lambda_{\beta\mu}.
\end{equation}
Using (\ref{connection}) we thus find
\begin{equation}
    \begin{split}
        R^\alpha{}_{\mu\beta\nu}=&\Big(\frac{a''}{a}-2\Big(\frac{a'}{a}\Big)^2\Big)\Big(\delta^\alpha_\nu\delta^0_\mu\delta^0_\beta-\delta^\alpha_0\delta^0_\nu\eta_{\mu\beta}-\delta^\alpha_\beta\delta^0_\nu\delta^0_\mu+\delta^0_\beta\delta^\alpha_0\eta_{\mu\nu}\Big)\\
        &-\Big(\frac{a'}{a}\Big)^2\Big(\delta^\alpha_\nu\eta_{\mu\beta}-\delta^\alpha_\beta\eta_{\mu\nu}\Big)\\
        R_{\mu\nu}=&R^\alpha{}_{\mu\alpha\nu}\\
        =&\Big(\frac{a''}{a}-2\Big(\frac{a'}{a}\Big)^2\Big)\Big(\eta_{\mu\nu}-(D-2)\delta_\mu^0\delta_\nu^0\Big)+\Big(\frac{a'}{a}\Big)^2(D-1)\eta_{\mu\nu}\\
        R=&R^\mu{}_\mu\\
        =&\Big(\frac{a''}{a}-2\Big(\frac{a'}{a}\Big)^2\Big)2(D-1)+\Big(\frac{a'}{a}\Big)^2D(D-1)
    \end{split}
\end{equation}

\end{document}